\documentclass[12pt]{revtex4-1}
\usepackage{amsmath}
\usepackage{amssymb}
\usepackage{comment}
\usepackage{graphicx}
\usepackage{xcolor}
\usepackage{cancel}
\usepackage[normalem]{ulem}
\usepackage{hyperref}

\usepackage{color}
\usepackage{soul}
\usepackage{url}
\usepackage{svg}
\usepackage[normalem]{ulem}
\usepackage{soul}
\sethlcolor{yellow} 

\begin{document}

\title{The wealth-to-salary ratio: a simple indicator of inequality 
}

\author{Jhordan Silveira de Borba $^{1}$}
\thanks{sbjhordan@gmail.com} 

\author{Sebastian Gonçalves $^{1}$}
\thanks{sebastiangoncalves@gmail.com}

\author{Celia Anteneodo $^{2}$}
\thanks{celia.fis@puc-rio.br}

\affiliation{$^{1}$Instituto de Física, Universidade Federal do Rio Grande do Sul, Porto Alegre/RS, Brazil \\ 
$^{2}$Departamento de Física, PUC-Rio, Rio de Janeiro/RJ, Brazil
}

\begin{abstract} 

In previous studies, using agent-based simulations of a model of a capitalist economy proposed by Ian Wright, we showed that inequality in both wealth and income distributions is largely governed by a single key parameter: the ratio between average wealth per capita and the mean salary, denoted by $R$. This parameter captures the relative importance of accumulated assets compared with  labor income. In the model, larger values of $R$ are systematically associated with higher levels of inequality, as measured by the Gini coefficient.
In the present work, we investigate whether this relationship is also observed in real-world economies. Using empirical data from the World Inequality Database (WID) and the Organization for Economic Co-operation and Development (OECD), we construct empirical estimates of $R$ and compare them with standard measures of income and wealth inequality. We find a positive relation between $R$ and inequality across several empirical settings, with the strongest relationships observed for mature and advanced economies. These results provide empirical support for the role of the wealth-to-salary ratio as a simple macroscopic indicator associated with economic inequality, while also highlighting the limitations imposed by differences in data definitions and measurement across countries and periods.
 
\end{abstract}
\maketitle

 \section{Introduction} 
\label{sec:introduction}

Economic activity can be interpreted through the lens of complex systems, in which collective outcomes emerge from  interactions among heterogeneous agents~\cite{JUSUP20221,yakovenko2009,Farmer2009}.
Within econophysics, this perspective has proved particularly useful for understanding empirical patterns in income and wealth distributions~\cite{yakovenko2009,shaikh}.  
Agent-based models (ABMs), particularly those based on wealth-exchange mechanisms~\cite{chatterjee2007econophysics,Patriarca2010}, have been widely used to investigate how inequality emerges and evolves under different market rules, regulatory policies~\cite{CARDOSO2020,Cardoso2021,LIMA2022} and interaction structures~\cite{Kohlrausch2024}.
Among them, the Social Architecture of Capitalism (SA) model introduced by Ian Wright~\cite{WRIGHT2005589} provides a simple description of interactions between firms and workers, and reproduces important stylized features of wealth distributions, including a two-regime structure~\cite{dragulescu2001,yakovenko2023}.
In recent works~\cite{Borba2025,Borba2026}, we systematically explored the parameter space of the SA model and identified a quantity that largely controls the separation between the two wealth regimes.
Denoted by $R$, this quantity is defined as the ratio of average wealth per capita, $\overline{w}$, to the mean wage,  $\overline{p}$. 
Within the model, the overall shape of the wealth distribution---including the emergence of a gap between classes, increasing concentration of wealth, and the disappearance of an intermediate stratum at high values of $R$---is largely determined by this ratio, with only minor sensitivity to system size. In particular, inequality, as measured by the Gini coefficient~\cite{PLATAPEREZ201579}, increases monotonically with $R$. 
These results remain robust under several realistic modifications of the original model~\cite{Borba2026}.

These findings raise a natural empirical question: does the relationship between $R$ and inequality identified in the model also appear in real economies? The present work addresses this question using observational data. The causal mechanism connecting $R$ to inequality was established through controlled computational experiments in the SA model~\cite{Borba2026}. We therefore do not attempt to establish causality from observational data alone. Instead, our objective is more limited: to determine whether empirical patterns in real economies are consistent with the relationship predicted by the model.
 
To this end, we construct  empirical estimates of $R$, using measures of average employee compensation and net personal wealth, drawing on data from the World Inequality Database (WID)~\cite{WID} and  the Organisation for Economic Co-operation and Development (OECD)~\cite{OECD}.  
  In Section~\ref{sec:data}, we describe the data sources, the definitions adopted, and the procedures used to make the comparisons as consistent as possible. 
The empirical analysis is developed in Sec:~\ref{sec:real}, where 
we also connect $R$ to the wealth-to-income ratio, $\beta$, widely studied in the literature on wealth accumulation and inequality, particularly in the work of Piketty and coauthors~\cite{repec:oup:qjecon:v:129:y:2014:i:3:p:1255-1310}. 
Conclusions and final remarks are  presented in Sec.~\ref{sec:final}

\section{Data and methodology}
\label{sec:data}

Both databases used in our empirical analysis, the World Inequality Database (WID)~\cite{WID} and the Organization for Economic Co-operation and Development (OECD)~\cite{OECD}, provide internationally comparable measures of income, wealth, labor compensation, and inequality, but they differ in definitions, source data, and methodological procedures.

We use the WID data primarily for the analyses based on long time series and distributional measures, covering both groups of countries and individual countries. The OECD data are used for a complementary analysis. Although WID   incorporates data from  OECD and other official statistical sources, differences in definitions and methodological procedures may result in different estimates for some variables.  Using both databases therefore allows us to examine whether the observed relationship between $R$ and inequality is robust to differences in data sources and methodology.

\subsection{WID methodology}
\label{sec:WID}

We first construct a balanced panel dataset, in which each country is observed in every year from 1995 to 2019. The period begins in 1995 and ends in 2019 in order to avoid the effects of the COVID-19 pandemic. The countries included in the analysis, listed in Table~\ref{tab:tab22}, are those classified by the International Monetary Fund (IMF) as advanced economies---or as industrialized economies, in the terminology used at the time---in its 1995~\cite{imf1995weo} report, with the exception of Iceland, which was excluded due to lack of data.

Inequality is measured by the Gini coefficient of pre-tax national income and the Gini index of net personal wealth. To estimate $R$, we approximate $\overline{p}$ by average employee compensation and $\overline{w}$ by net personal wealth, both expressed per adult individual and in local currency.

According to the WID definitions, national income aims to measure the total income available to the residents of a given country~\cite{WIDglossario}. Net personal wealth corresponds to the total value of non-financial and financial assets held by households, minus their debts, while employee compensation corresponds to the total remuneration, in cash or in kind, paid by employers in return for work done.

For the aggregate analysis, we need to account for differences in population size across countries.  Otherwise, assigning equal weight to each country would give the same importance to smaller and more populous economies.  Therefore, we calculate the population-weighted averages of the Gini coefficient and the ratio $R$.

\begin{table}[t!]
\centering
\begin{tabular}{|c|l|c||c|l|c|}
Rank                   & Country        &  GDP  & Rank  & Country  & GDP   \\ \hline
1   & United States  &  32.4 & 12   & Ireland     & 0.779 \\
2   & Germany        &  5.45 & 13   & Belgium$^1$ & 0.777 \\
3   & Japan          & 4.38 & 14    & Sweden      & 0.760 \\
4   & United Kingdom & 4.26 & 15    & Austria$^*$ & 0.624 \\
5   & France         & 3.60 & 16    & Norway$^1$  & 0.599 \\
6   & Italy          & 2.74 & 17    & Denmark$^1$ & 0.504 \\
7   & Canada$^1$     & 2.51 & 18    & Portugal    & 0.381 \\
8   & Australia      & 2.12 & 19    & Finland     & 0.338 \\ 
9   & Spain$^*$      & 2.09 & 20& Greece$^1$  & 0.308 \\  
10  & Netherlands$^1$ & 1.45 & 21  & New Zealand$^2$ & 0.279 \\
11  & Switzerland    & 1.15 &  22   & Luxembourg$^*$  & 0.110 
\end{tabular}
\caption{Countries included in the analysis, ranked by nominal GDP at current prices (trillions of US dollars, 2026). For each country, the Spearman correlation coefficient between $R$ and the wealth and income Gini indices were calculated. All statistically significant coefficients were positive, except for those marked with superscripts: (1) one coefficient was negative; (2) both coefficients were negative; (*) neither coefficient was statistically significant.
}
\label{tab:tab22}
\end{table}

\subsection{OECD methodology}
\label{sec:OECD}

While the WID data allow us to examine the evolution of $R$ and inequality over time, the OECD data complement this analysis by providing an independent dataset for cross-country comparisons within a specified time window.

Ideally, cross-country comparisons would use observations from the same year. However, the available OECD data do not allow us to satisfy this condition for all countries. We therefore select a time window and, for each country, use the most recent available observation within that window for which all the variables required for the analysis are available.

We choose the 2015--2019 period, balancing  the need to limit temporal heterogeneity against the need for sufficient data coverage to include a reasonable number of countries, while excluding the years affected by the COVID-19 pandemic. 
We restrict the analysis to countries classified by the IMF as advanced economies since at least the 2000s, focusing on economies with a long history of industrialization~\cite{BackMatter}. 
For each country,, we require $\overline{w}$, $\overline{p}$, and the Gini coefficient to refer to the same year. This criterion yields 17 countries with all the necessary data available, listed in Table~\ref{tab:OECD}.

\begin{table}
\centering
\begin{tabular}{l|l||l|l}
Country & Year & Country & Year \\ \hline
Australia & 2018 & Luxembourg & 2018 \\
Austria & 2017 & Netherlands & 2019 \\
Canada & 2019 & New Zealand & 2018 \\
Denmark & 2019 & Norway & 2018 \\
Finland & 2019 & Portugal & 2017 \\
Germany & 2017 & Spain & 2018 \\
Greece & 2018 & United Kingdom & 2017 \\
Ireland & 2018 & United States & 2019 \\
Italy & 2016 & & \\
\end{tabular}
\caption{Countries included in the analysis using OECD data, listed in alphabetical order, together with the year of the data used. All countries were classified as advanced economies by the IMF in the 2000s.}
\label{tab:OECD}
\end{table}

In the OECD data analysis, $\overline{w}$ is approximated by net wealth per person, measured at current prices, while $\overline{p}$ is approximated by the average annual wage, also expressed at current prices, everything in the local currency. Inequality is measured using the Gini coefficient of disposable income.

It is worth-noting that, according to the OECD definition, wealth refers to the ``ownership of economic capital'' as a dimension of people's material well-being. Average annual wages per full-time equivalent dependent employee are obtained by dividing the national-accounts-based total wage bill by the average number of employees in the total economy and converting this quantity into full-time equivalent units using the ratio of average usual weekly hours per full-time employee to those of all employees. The Gini coefficient is based on the comparison of cumulative proportions of the population against cumulative proportions of the income they receive.

\section{Empirical analysis of $R$ using real data}
\label{sec:real}

In this Section we examine whether the positive relationship between $R$ and inequality identified in the SA model is also observed in empirical data. We proceed from an aggregate to more  disaggregated analyses. 

(i) First, in Section~\ref{sec:group}, we analyze advanced economies in groups, using WID   and OECD data.

(ii) Before proceeding to individual analysis, we briefly discuss, in Section~\ref{sec:piketty}, the connection between $R$ and the wealth-to-income ratio, $\beta$, widely used in the literature following Piketty. This comparison helps clarify the relation between the two indicators and provides a broader context for the empirical analysis.  

(iii) Finally, we examine, in Section~\ref{sec:single}, individual countries using longer WID time series. 
 
 This progression allows us to assess the robustness of the relationship across different aggregation levels, data sources, and historical settings. 

To assess the strength and direction of the association between different quantities, we compute the Spearman rank correlation coefficient, $\rho_s$, a nonparametric measure of monotonic dependence. Throughout the text, statistical analyses are evaluated at a significance threshold of $p \leq 0.05$ and for Spearman  correlations, statistical significance is assessed using a permutation test. 

\newpage
\subsection{Advanced economies}
\label{sec:group}
\subsubsection{Aggregate analysis}
\label{sec:aggregate}

 \begin{figure}[b!] 
 \centering
    \includegraphics[width = 0.49\textwidth]{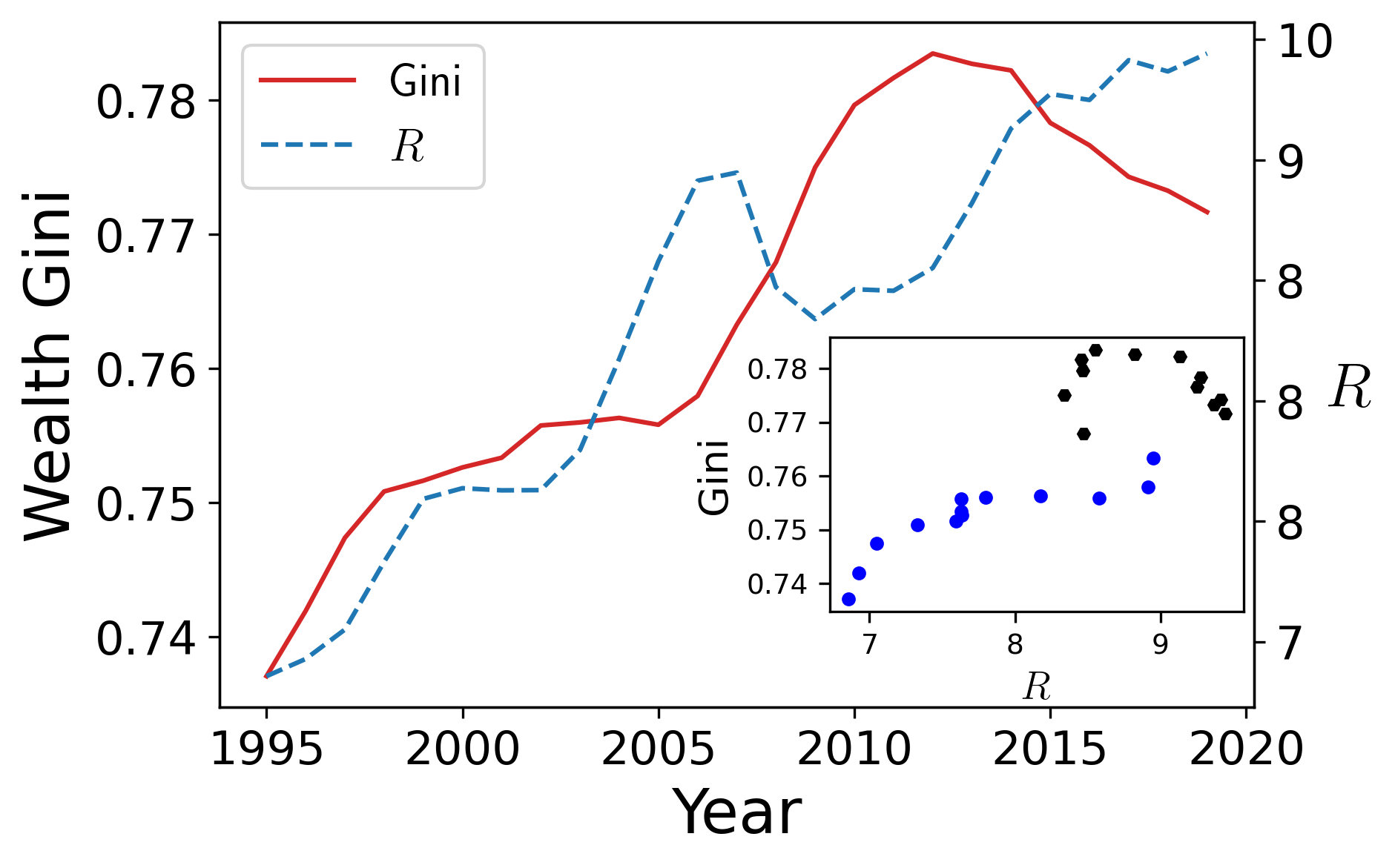}
\includegraphics[width = 0.49\textwidth]{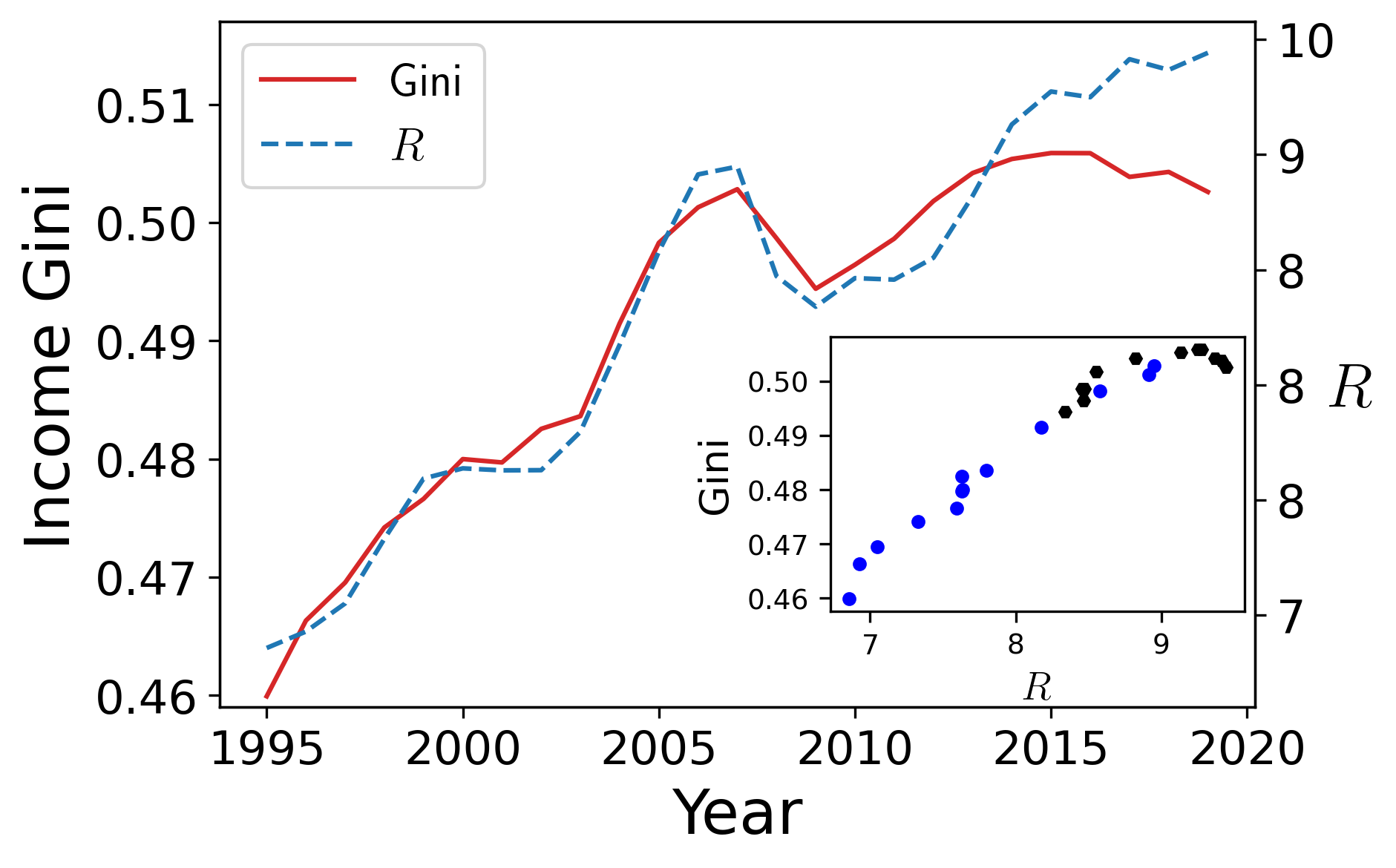}
\caption{Time series of $R$ and the wealth and income Gini coefficients, calculated as population-weighted  annual averages across the 22 selected advanced  economies over the 25-year period (1995–2019).  The insets show the corresponding Gini  coefficients plotted against $R$.  
Light-blue symbols represent the period 1995--2007, while black symbols represent 2008 onward. The ratios of the standard deviations to the corresponding means remain relatively stable over the period, ranging from $0.13$ to $0.25$ for $R$, from $0.08$ to $0.10$ for the wealth Gini coefficient, and from $0.11$ to $0.13$ for the income Gini coefficient.
}
 \label{fig:22countries}
\end{figure}

We aggregate the data for the advanced economies  of Table~\ref{tab:tab22}, as described in Sec.~\ref{sec:WID}, in order to examine whether the relationship between $R$ and inequality is visible at the level of the combined population.  The relationships between the wealth and income Gini coefficients and $R$ are shown in Fig.~\ref{fig:22countries}, with Spearman correlation coefficients of $\rho_s=0.73$ and $\rho_s=0.95$, respectively.
Variations in $R$ tend to be accompanied by changes in income inequality (right panel), even after the 2008 financial crisis, when the relationship between the wealth Gini coefficient and $R$ becomes more complex (left panel).

For wealth inequality, we also estimated the Gini coefficient for the pooled population of the 22 countries, using the population size and per capita wealth of each decile in each country, with wealth converted from local currencies to PPP dollars. Although this procedure involves several approximations, the resulting time series showed qualitatively similar behavior to the average Gini values presented in Fig.~\ref{fig:22countries}. However, the decile-based estimates were consistently lower than the average of the country-level coefficients.
Since decile-level income data are not available, we cannot estimate the income Gini coefficient for the pooled population using the same procedure used for the wealth Gini coefficient.

 To determine whether the aggregate relationship is also reflected at the country level, we next apply the Spearman  correlation test separately to each of the 22 countries. We retain only statistically significant correlations.  Table~\ref{tab:spearman} summarizes the results.

\begin{table}[h!]
\centering
\begin{tabular}{c|c|c}
         & Wealth & Income \\
         \hline
Positive & 13& 13\\
Negative & 5      & 3      \\
\hline
Total    & 18& 16\end{tabular}

\caption{Summary of the signs of statistically significant Spearman correlations between $R$ and the  wealth and income Gini coefficients across the 22 countries considered.  The entries indicate the number of countries with positive or negative correlations for each inequality measure.
}
\label{tab:spearman}
\end{table}

As shown in Table~\ref{tab:tab22}, 19 of the 22 countries yielded at least one statistically significant result. 
Moreover, 15 countries yielded significant results for both measures and 63\% of the countries with at least one statistically significant correlation showed no  negative correlations. This group includes six of the world’s top ten largest by GDP in 2026, while the remaining four lie outside this study~\cite{imfgpd}. As shown in Table~\ref{tab:tab22},  the first half of the ranking is dominated by countries with at least one positive correlation, whereas negative correlations become more common in the second half.  

\subsubsection{Cross-country analysis}
\label{sec:individual}

We next turn to the cross-country analysis based on OECD data to examine whether the positive association between $R$ and inequality observed in the aggregate WID analysis also holds in a cross-sectional comparison using a different dataset. We consider the countries listed in Table~\ref{tab:OECD}, all of which are OECD members and were classified by the IMF as advanced economies in the 2000s.

\begin{figure}[h!] 
     \centering
    \includegraphics[width = 0.6\textwidth]{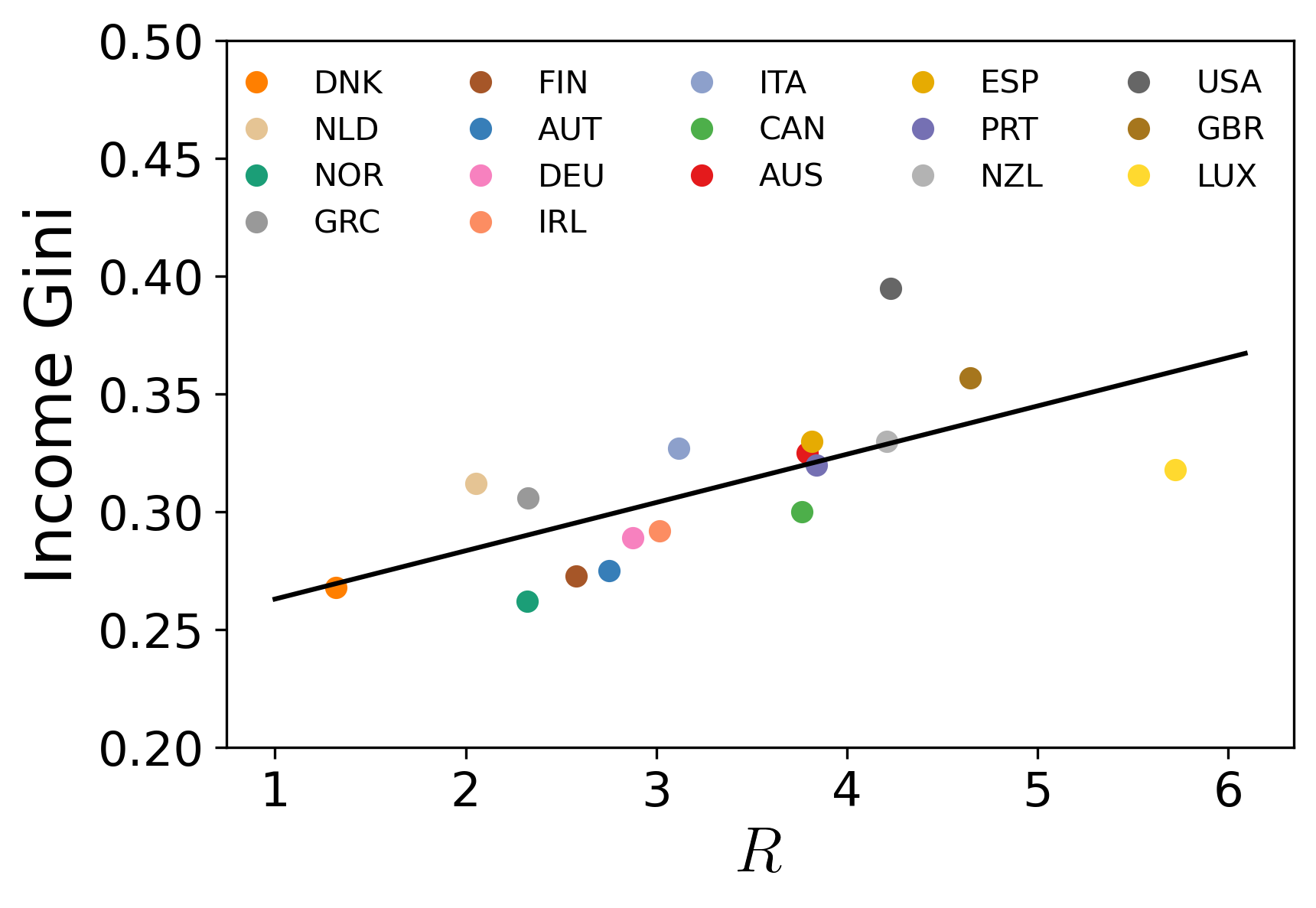}
    \caption{Income Gini coefficient versus the $R$ ratio, based on OECD data for the countries and corresponding years listed in Table~\ref{tab:OECD}, covering the period 2015--2019.}
    \label{fig:R-OECD}
\end{figure}

In Fig.~\ref{fig:R-OECD}, we present the income Gini vs $R$, for the countries in Table~\ref{tab:OECD} where the year considered for each country is specified.

Despite the challenges involved in cross-country comparisons, the data show a moderate positive association between inequality and $R$, with correlation  $\rho_S=0.78$. 

 \subsection{Comparison of $R$ with the wealth-to-income ratio $\beta$}
 \label{sec:piketty}
 
To place our indicator in the context of the existing literature, we compare $R$ with the wealth-to-income ratio. In Piketty's terminology~\cite{piketty}, the latter is denoted by $\beta$ and measures the ratio of aggregate wealth to national income. Although $\beta$ is closely related to $R$, the relationship between inequality and the wealth-to-wage ratio considered here has not, to our knowledge, been specifically investigated.

In a study by Piketty and Zucman~\cite{repec:oup:qjecon:v:129:y:2014:i:3:p:1255-1310}, eight advanced economies are analyzed, and the authors concluded that in all these countries there has been a gradual increase in the wealth-income ratio in recent decades.  According to the authors, this increase is not necessarily problematic in itself, but it raises important questions concerning taxation and regulation, since high wealth-to-income ratios imply that wealth inequality is likely to play a central role in the overall structure of inequality in the twenty-first century.

Madsen, on the other hand, focuses on the last eight centuries of British history and engages in critical dialogue with Piketty's work, but also concludes that the data suggest that variations in income inequality are accompanied by variations in $\beta$, i.e., the time series for income inequality follows a pattern very similar to that of $\beta$. The author further suggests that the wealth-to-income ratio may serve as a proxy for inequality~\cite{MADSEN2019246}. 

Given the documented relationship between inequality and $\beta$, it is important to clarify how $\beta$ and $R$ are related. The two ratios differ in their denominators: $\beta$ uses national income, whereas $R$ uses mean wages. Their numerators are also expressed differently, with $\beta$ using national wealth and $R$ using per capita wealth. However, this distinction is purely a matter of normalization, since the ratio of per capita wealth to per capita income is identical to the ratio of national wealth to national income. While an increase in national income does not necessarily translate into a proportional increase in mean wages, the two quantities are structurally related because labor compensation is a component of national income. Thus, changes in national income are expected to be associated with changes in average wages~\cite{Przekota2023CausalityIT}.

Consequently, $\beta$ and $R$ may exhibit a similar qualitative long-run behavior insofar as average wages tend to increase with national income. The two ratios need not, however, evolve proportionally. Their relative dynamics depend, in particular, on the labor share, {\em i.e.}, the fraction of national income paid to workers as compensation.

To compare $\beta$ and $R$, we first specify how each quantity is estimated from the empirical data.
We use average compensation per capita for the total population, $\overline{c}$, reported by WID, as a proxy for the average wage received by workers, $\overline{p}$, in our model~\footnote{Wages and compensation differ in definition: our model uses wages, whereas WID reports compensation, which includes wages and social security contributions.}.  
We also obtain national income, $Y$, and total net personal wealth, $W$, from WID. 
Denoting the labor share of national income by $s$ and the total population by $N$, we write, as a first approximation, $\overline{p} \approx \overline{c} = sY/N$. Likewise, we use net personal wealth per capita as a proxy for average wealth, $\overline{w} \approx W/N$. Under these approximations, $R$ can be expressed as 
  
\begin{equation} \label{eq:R-beta}
   R \approx \frac{\overline{w}}{\overline{p}} = \frac{W/N}{s Y/N}=\frac{1}{s}\frac{W}{Y} \approx \frac{1}{s}\beta. 
\end{equation}
In other words, $R$ is proportional to $\beta$ if the labor share $s$ remains constant.   

\begin{figure}[t!]
    \centering

    \includegraphics[width=0.49\linewidth]{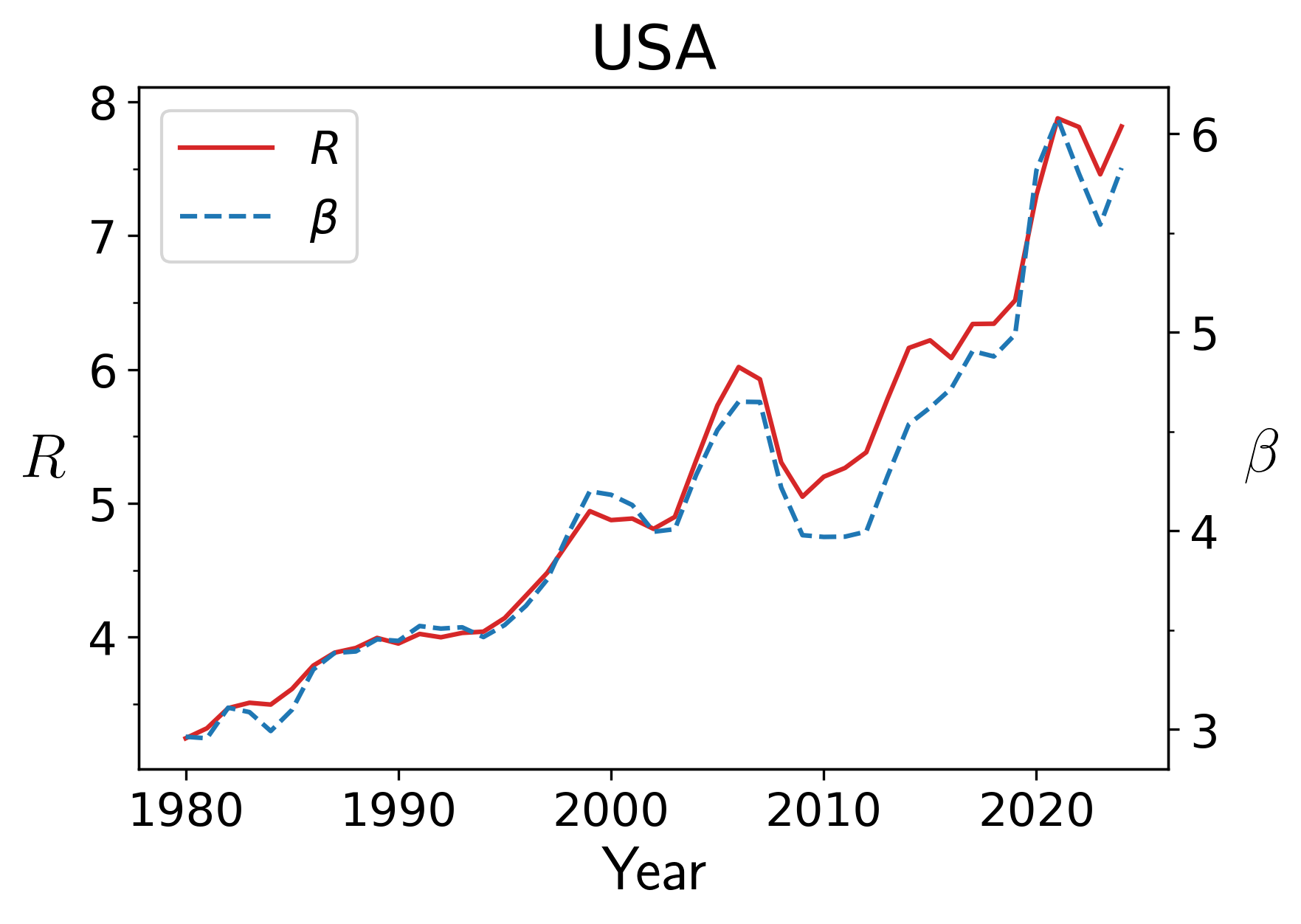}
    \includegraphics[width=0.49\linewidth]{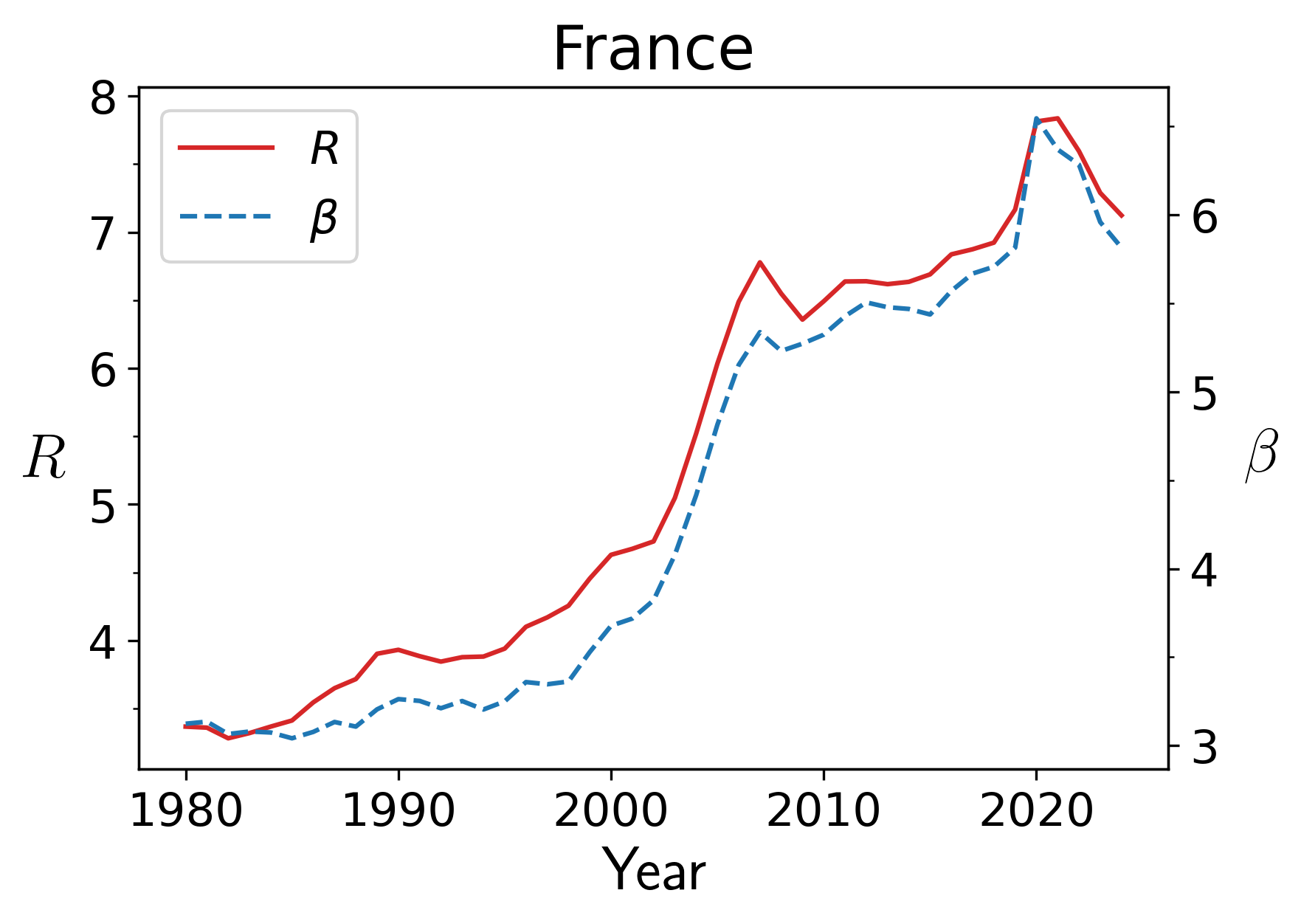}
    \includegraphics[width=0.49\linewidth]{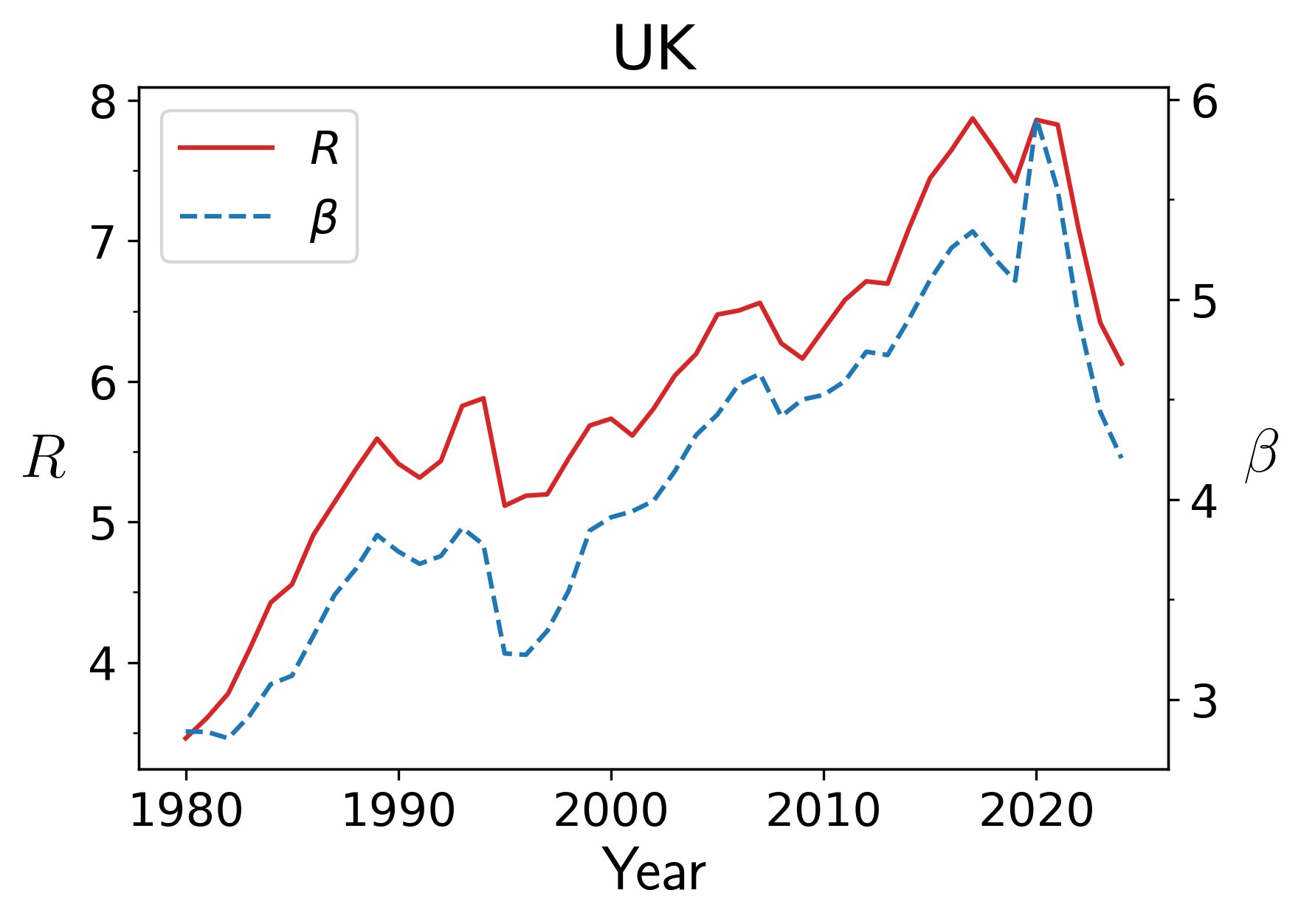}
    \includegraphics[width=0.49\linewidth]{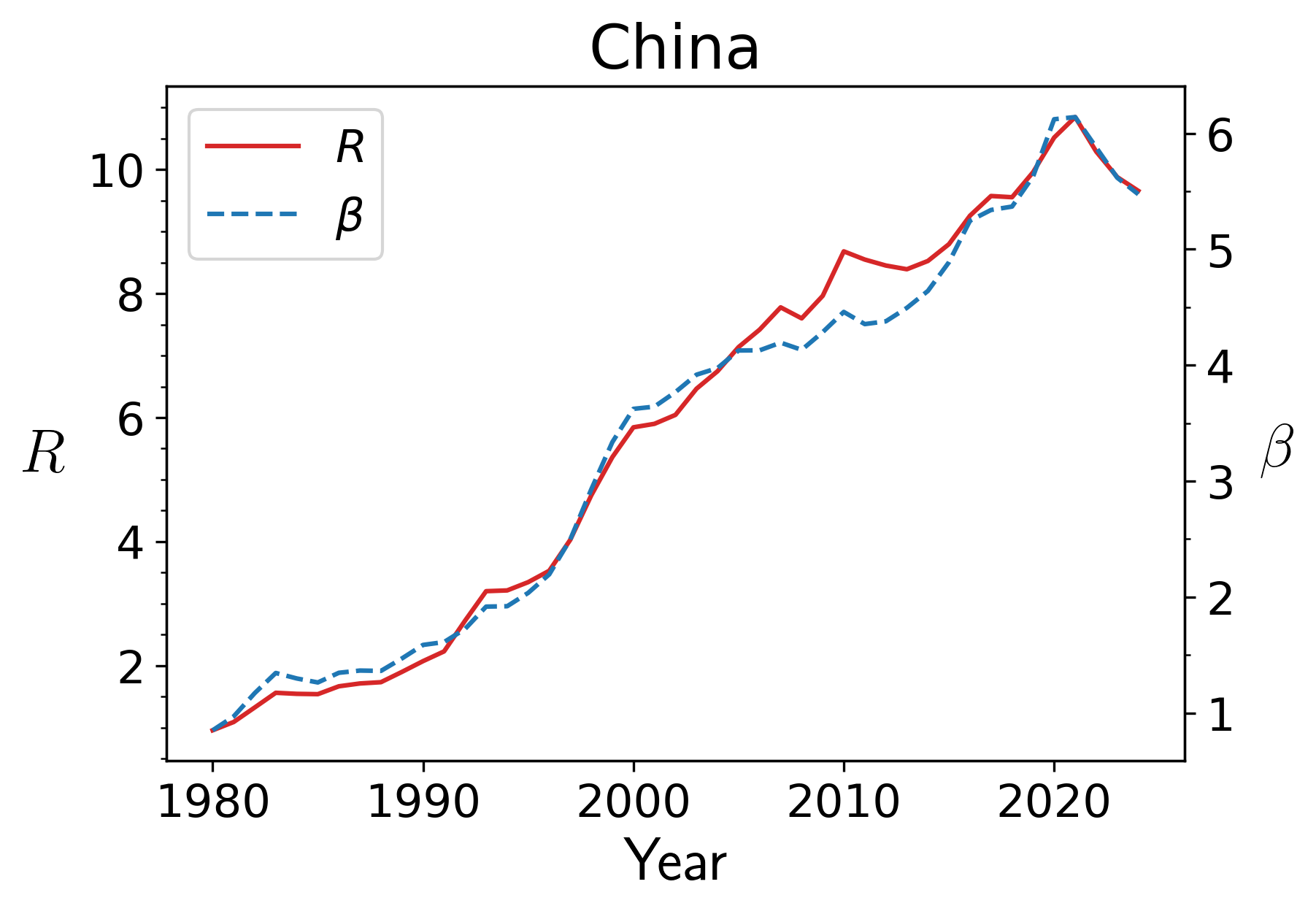}
    \caption{Comparison of the time series of $R$ and $\beta$ for the USA, France, the UK, and China over the period 1980-2024.}
    \label{fig:comparison}
\end{figure}

\begin{figure}[b!]
    \centering
    \includegraphics[width=0.6\linewidth]{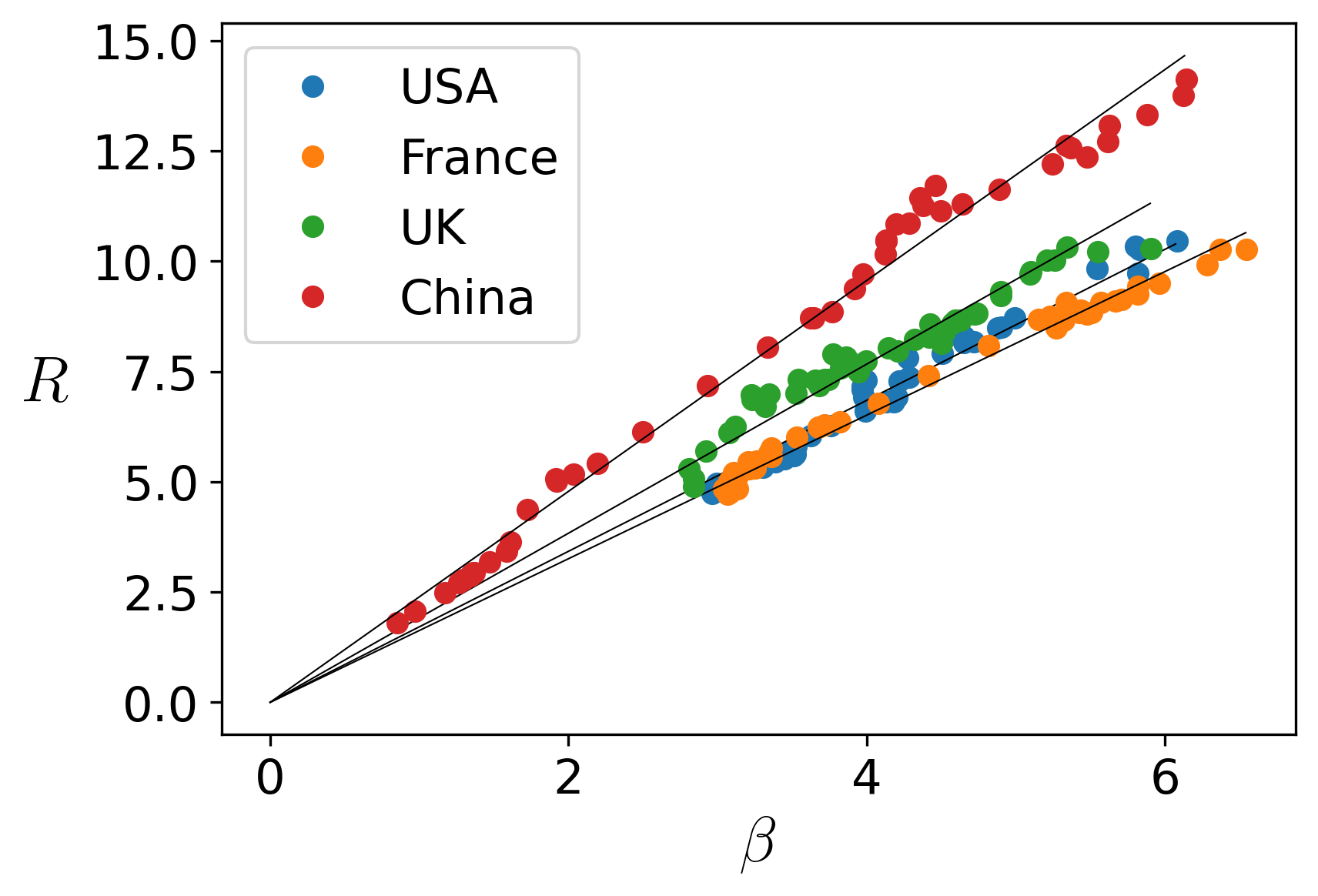}
    \caption{Scatter plot of $R$ versus $\beta$ for the countries indicated in the legend. Each symbol represents one annual observation over the period 1980–2024.}
    \label{fig:comparison2}
\end{figure}

The comparison of the time series for  $R$ and $\beta$ over the period 1980-2024 is shown in Fig.~\ref{fig:comparison}, revealing that they follow very similar patterns not only in their overall long-term trends but also in their shorter-term fluctuations. Additional information is provided in Fig.~\ref{fig:comparison2},  where a nearly linear relationship for all the four countries emerges, as expected from Eq.~(\ref{eq:R-beta}).  
Slope $\alpha$ of the fitted line constrained to pass through the origin together with the corresponding average value of $1/s$, are shown in Table~\ref{tab:slops}, for the USA, France, the United Kingdom, and China. The ratio of the fitted slope to the average inverse labor share ranges from approximately $0.99$ to $1.01$, supporting the consistency of the approximation $sR \approx \beta$ with the empirical results.

\begin{table}[h!]
    \centering
    \begin{tabular}{l|c|c}
       Country  &   $\alpha$&   $1/s$\\
       \hline
    USA  &  $1.71 \pm 0.01$&  $1.69 \pm 0.07$\\
    France  &  $1.63 \pm 0.01$&   $1.63 \pm 0.05$\\
    UK  &  $1.92 \pm 0.01$&   $1.93 \pm 0.09$\\
    China  &   $2.39 \pm 0.02$&   $2.36 \pm 0.16$\end{tabular}
    \caption{Slope of the linear fit constrained to pass through the origin ($\alpha$) and the inverse of the mean labor share ($1/s$), calculated from the same data series.}
    \label{tab:slops}
\end{table}

\subsection{A closer look to individual countries}
\label{sec:single}
 
\begin{figure}[t!] 
     \centering
    \includegraphics[width = 0.43\textwidth]{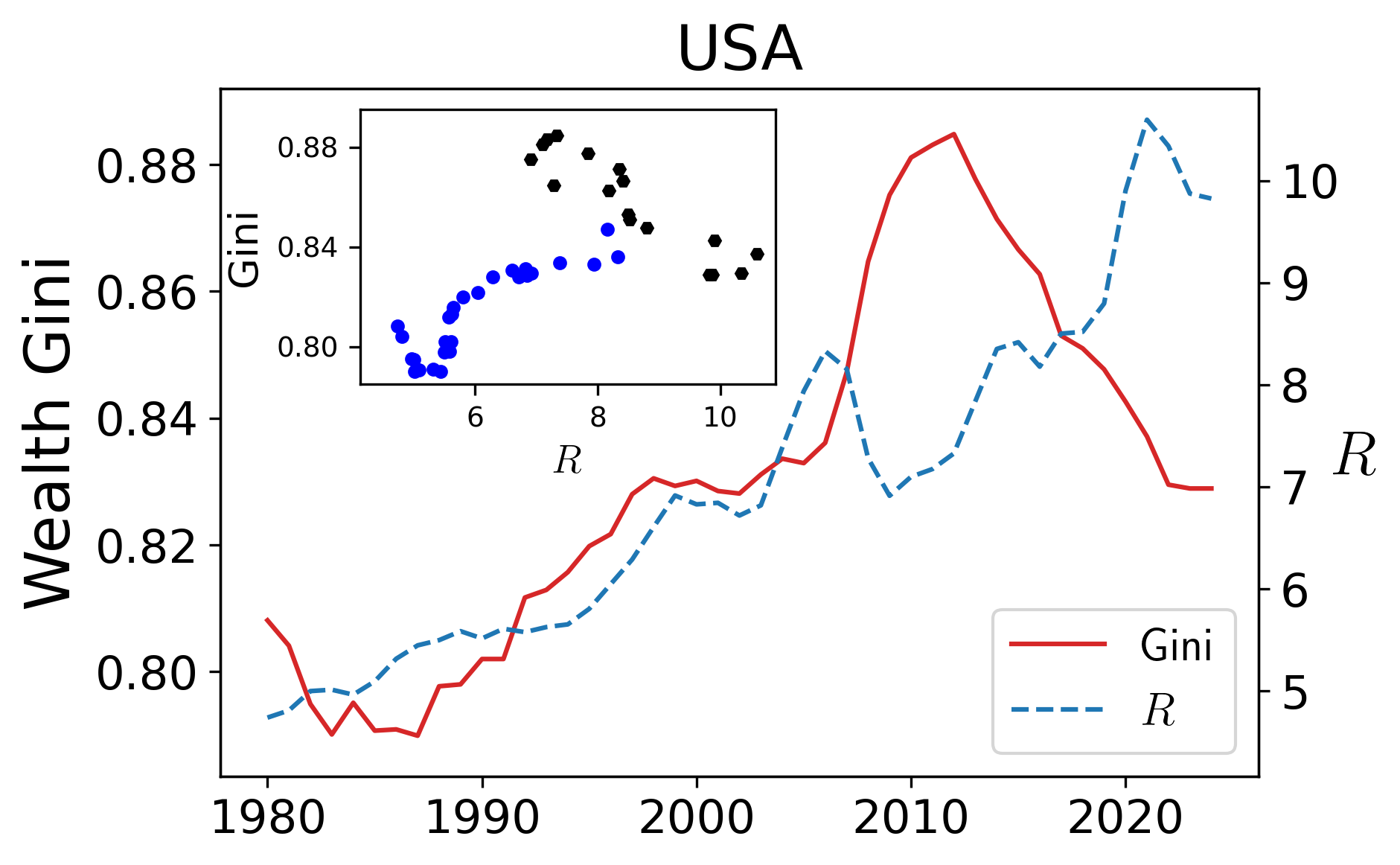}
    \includegraphics[width = 0.43\textwidth]{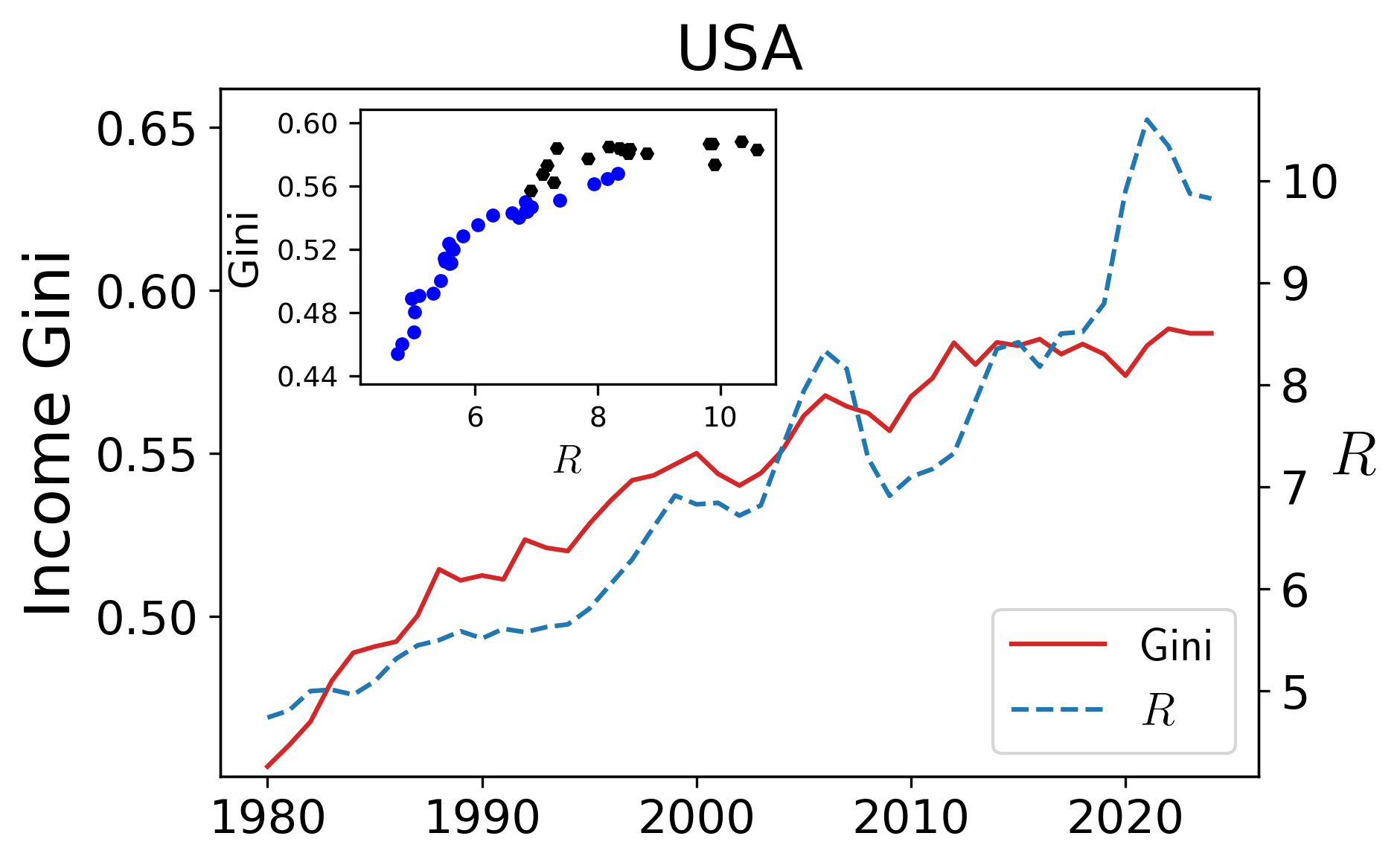}
    \includegraphics[width = 0.43\textwidth]{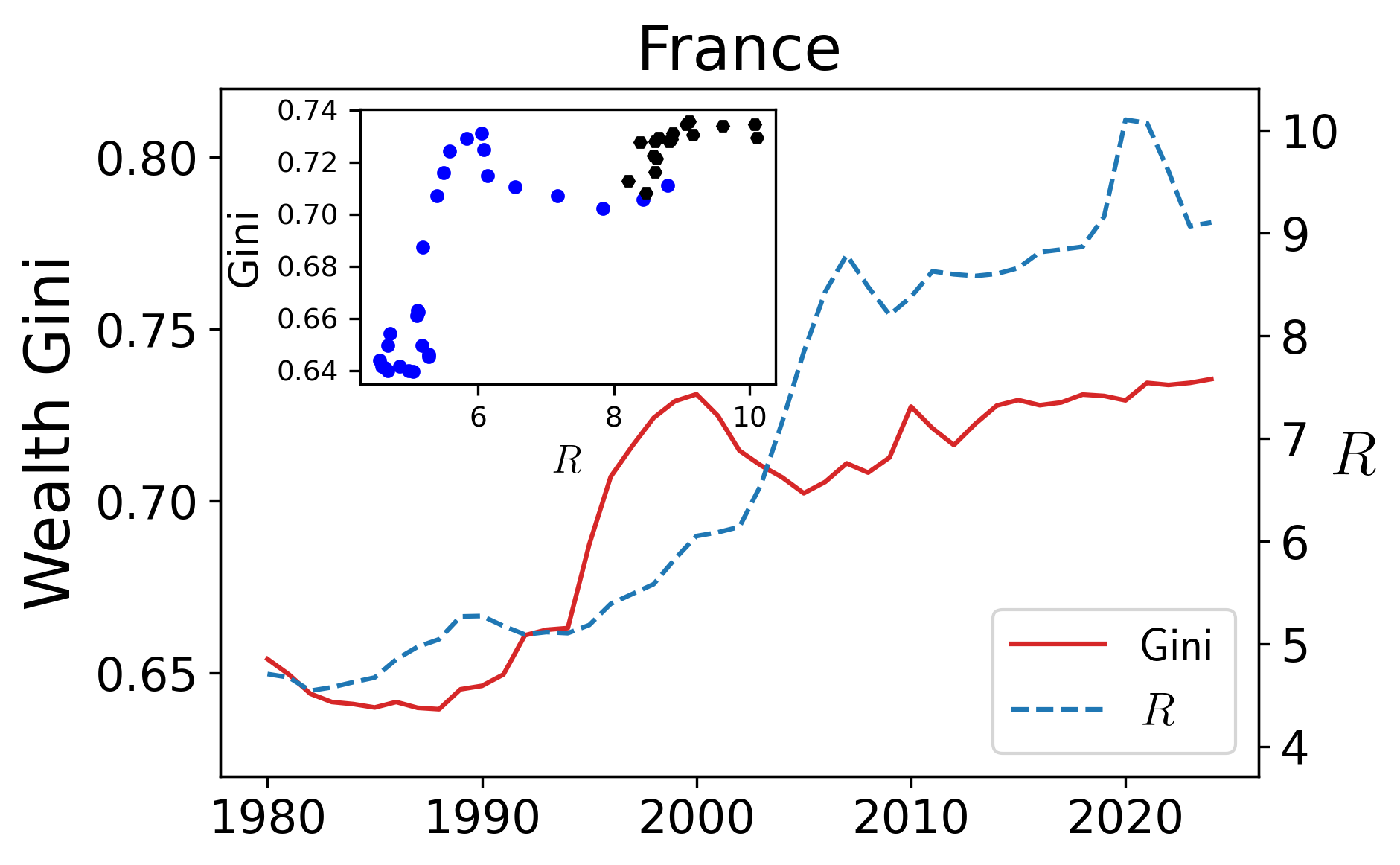}
    \includegraphics[width = 0.43\textwidth]{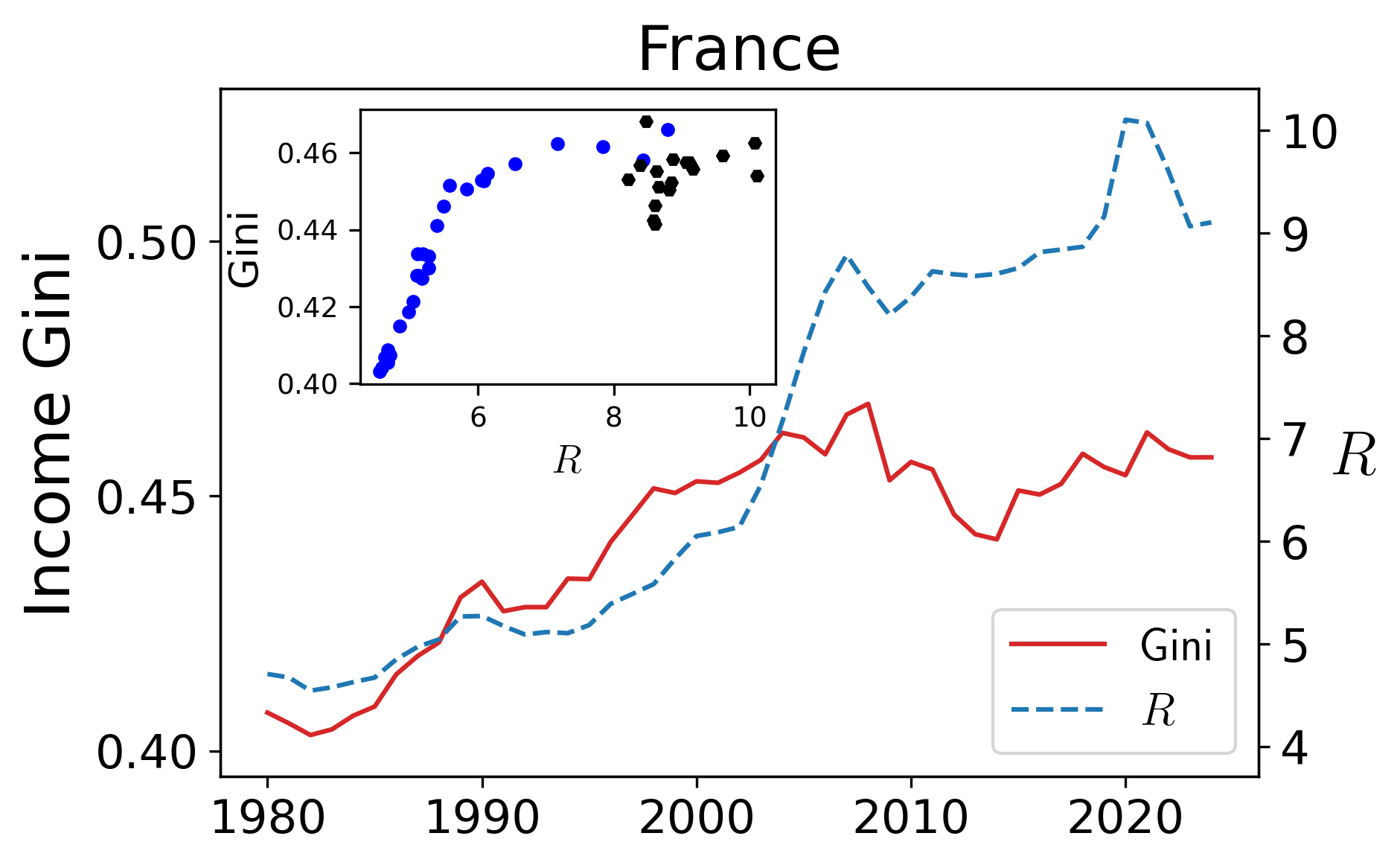}
    \includegraphics[width = 0.43\textwidth]{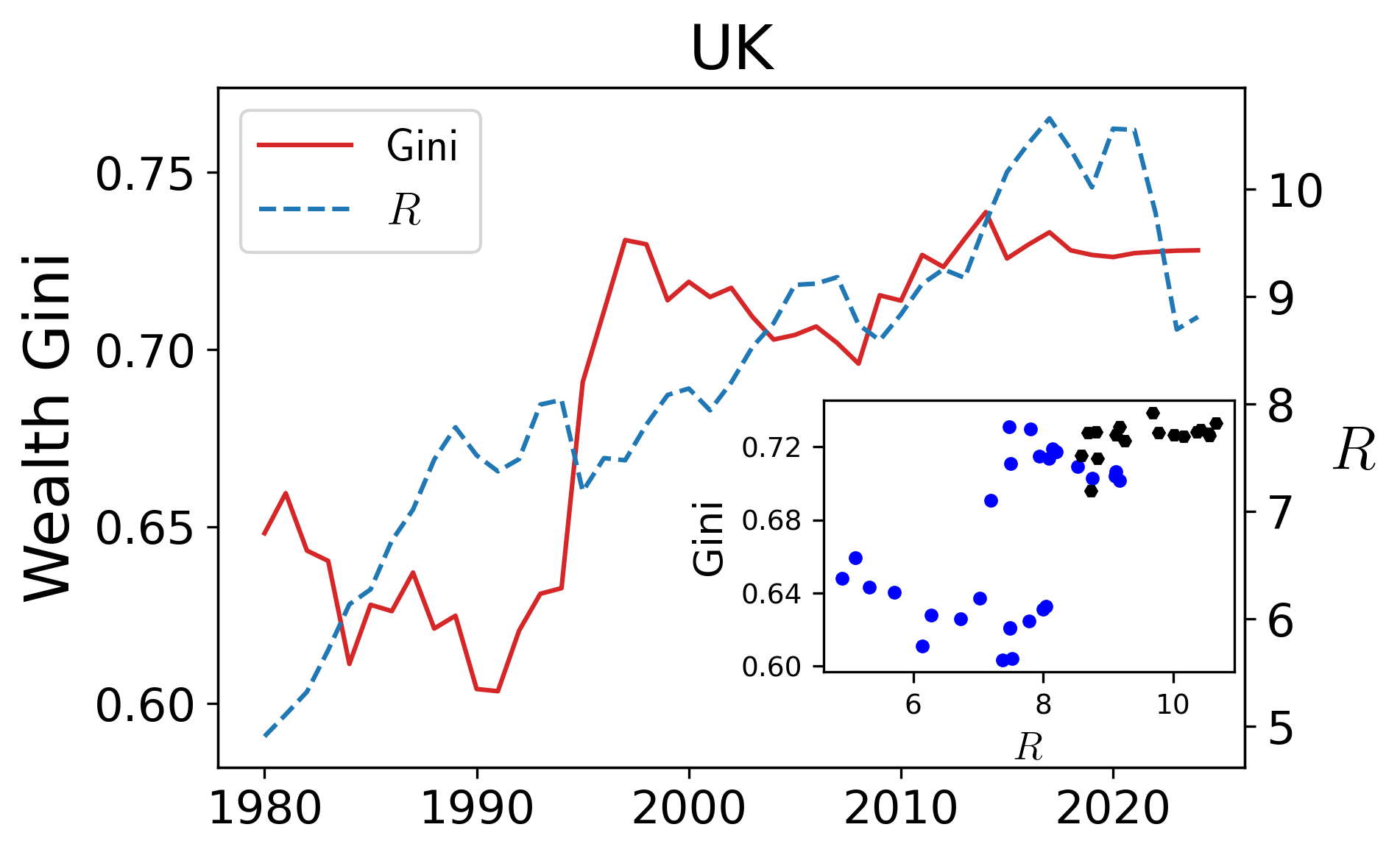}
    \includegraphics[width = 0.43\textwidth]{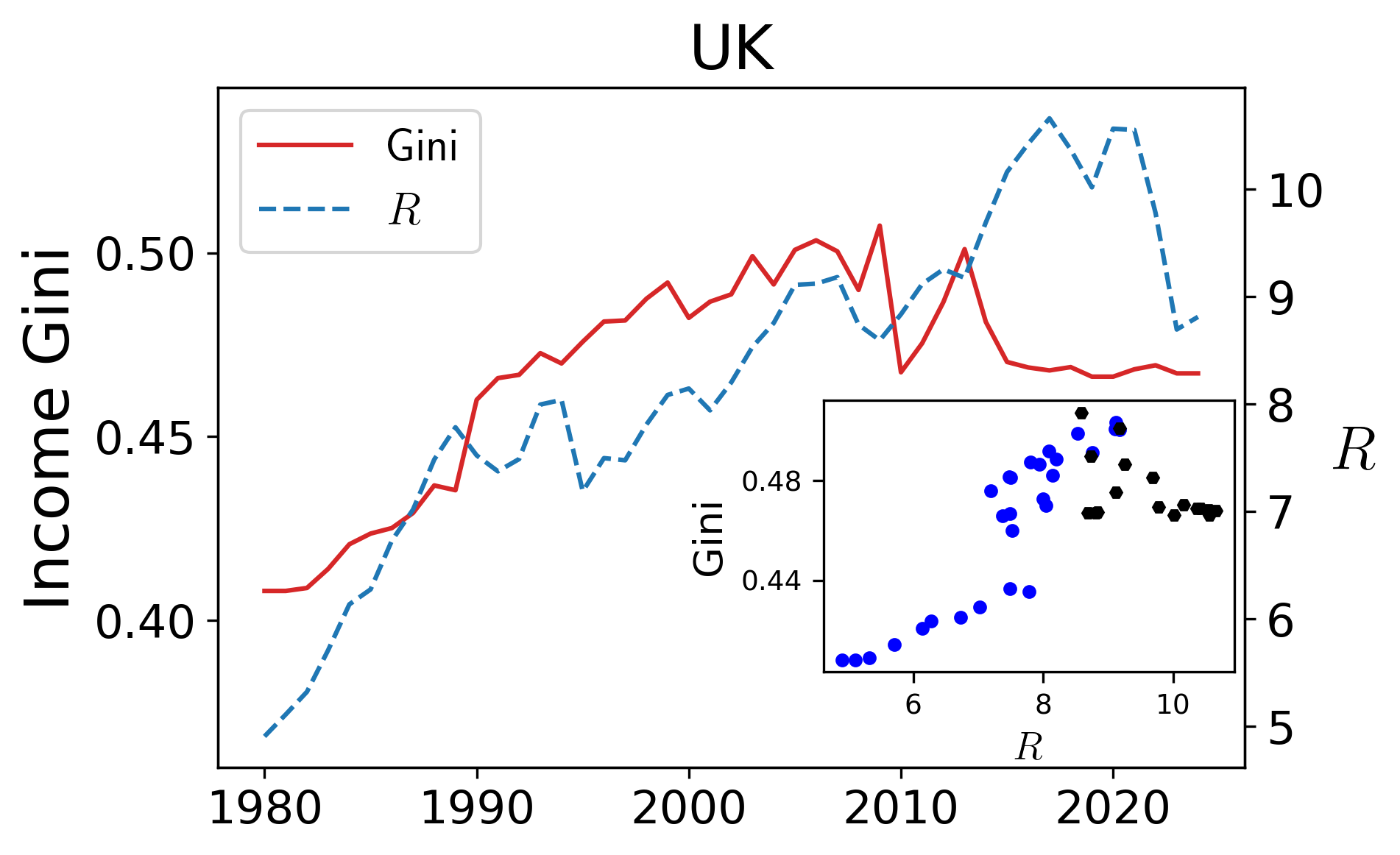}
    \includegraphics[width = 0.43\textwidth]{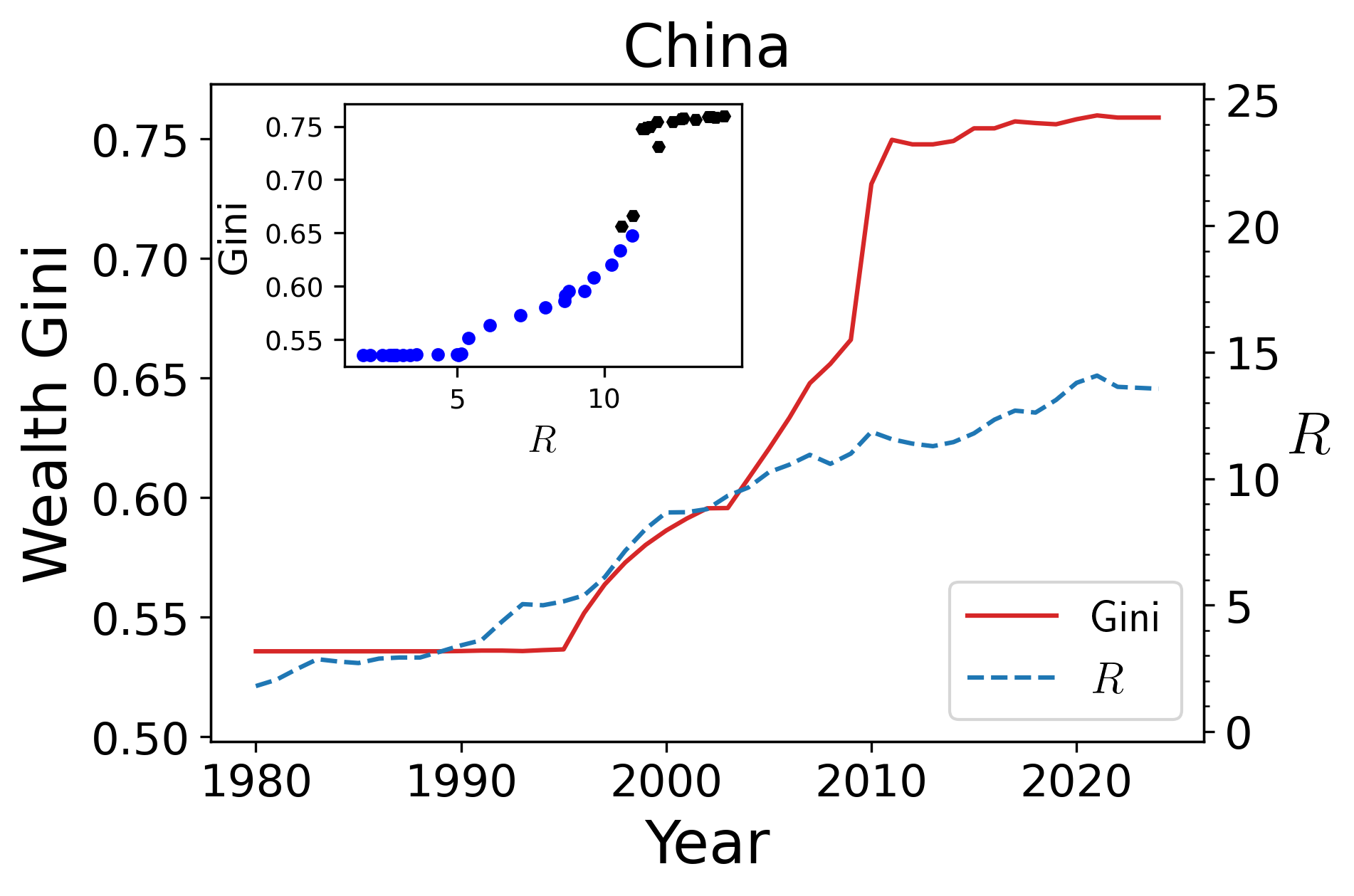}
    \includegraphics[width = 0.43\textwidth]{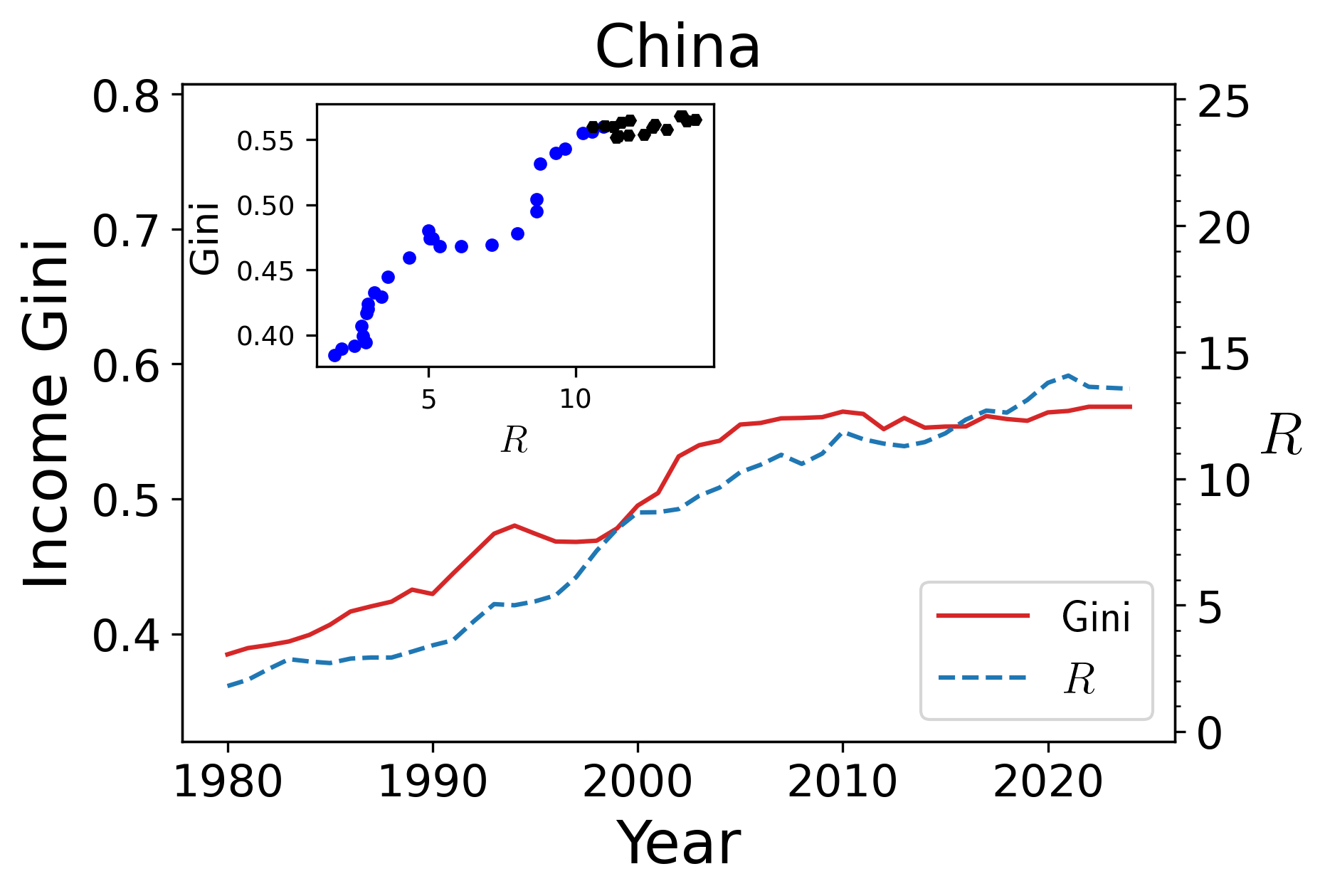}
    \caption{Time series of $R$ and the wealth (first column) and income (second column) Gini coefficients from 1980 to 2024 for the USA, France, the UK, and China, with one country per row. The insets show the corresponding Gini coefficient plotted against $R$. Light-blue symbols represent the period 1980–2007, while black symbols represent 2008 onward.
}
 \label{fig:all4}
\end{figure}

We now consider the time series of $R$, from 1980 to 2024, together with the income and wealth Gini indices for the USA, France, the UK, and China. All data are obtained from the World Inequality Database (WID)~\cite{WID}. The timeseries are summarized in Fig.~\ref{fig:all4}.
The insets represent the Gini-measured inequality vs. $R$,  discriminating between the periods before (light blue) and after (dark black) the 2007 Global Financial Crisis.

As in the aggregate analysis, the correlation between $R$ and income inequality is stronger and more statistically significant than that between $R$ and wealth inequality, particularly before the 2008 financial crisis, as shown in Table~\ref{tab:correlations}.
\begin{table}[h!]
    \centering
    \begin{tabular}{l|c|c}
       Country  &   Wealth &   Income    \\
       \hline
    USA     &  0.77 &  0.96\\
    France  &  0.83 &  0.87  \\
    UK      &  0.47 &  0.70  \\
    China   &  0.96 &  0.99   
    \end{tabular}
    \caption{Spearman correlation coefficients between $R$ and the wealth and income Gini coefficients for each country.}
    \label{tab:correlations}
\end{table}

Therefore, it is worth noting that, although the crisis is typically dated from 2008 to 2009, its effects persisted for several years. For instance, in the USA, as late as September 2015, full-time employment had not yet returned to its pre-crisis level~\cite{emprego-recessao}.  In absolute terms, the November 2007 level was only surpassed in October 2015. Similarly, full-time employment in January 2022 had not yet recovered to its pre-pandemic level of 2019.  The December 2019 level had not yet been exceeded by January 2022.  This indicates that the concurrent increase in $R$ and the Gini coefficient, whether for wealth or income inequality, is not confined to periods of crisis. Rather, the results suggest that this pattern is more characteristic of non-crisis periods.

Piketty argues~\cite{piketty} that the tendency toward increasing inequality under capitalism was temporarily reversed between the two world wars and during the subsequent decades, a period marked by the expansion of the welfare state and the implementation of fiscal policies such as progressive taxation. According to this interpretation, the 1980s marked a return to policies associated with market deregulation and a renewed increase in inequality~\cite{piketty}. The rise in inequality since the 1980s is broadly observed across the countries in our dataset and is accompanied by an increase in $R$ over the same period.

Because China's transition from a centrally planned economy to a market-oriented system began only after the economic reforms initiated by Deng Xiaoping in 1978, consistent measures of income and wealth inequality are available only from the early 1980s onward. The reforms gradually introduced market mechanisms, private entrepreneurship, and foreign investment while maintaining significant state control over strategic sectors. Consequently, the Chinese observations in our dataset start in 1980, corresponding to the onset of the country's transition toward a socialist market economy and the emergence of meaningful variation in both income and wealth inequality.

While France and the UK are especially valuable because of their long historical series of economic data, and the USA is of particular historical interest given its emergence as a leading capitalist power in the postwar period, the inclusion of China is especially informative because it provides a long-term record of inequality dynamics during a rapid transition from a predominantly egalitarian planned economy to a market-oriented system. The Chinese case therefore offers a valuable opportunity to examine how the relationship between income Gini index and the wealth-to-wage ratio ($R$) evolves during a period of profound structural and institutional transformation.

Despite some quantitative differences across countries, these empirical results suggest that the positive association between $R$ and inequality predicted by the SA model is also apparent in the long-term evolution of several individual countries whose economies are characterized by stable mass production of goods for market exchange.

\section{Final remarks and conclusions}
\label{sec:final}

The results suggest that the association between $R$ and inequality predicted by the SA model is particularly evident in advanced, long-established economies. In this group, the association with income inequality is statistically significant. This finding motivates a discussion of how the model’s assumptions relate to the institutional characteristics of the economies analyzed.

According to Erik Olin Wright’s conception of capitalism, describing an economy as capitalist means that capitalist relations are dominant within its economic ecosystem, without necessarily being the only relations present~\cite{wright2022ser}. Since the SA model represents an idealized capitalist system, we may hypothesize that its predictions are more readily observed in economies where these relations are firmly institutionalized. China provides a complementary case for examining this hypothesis: its transition toward a more market-oriented economy involves structural and institutional changes that may affect both $R$ and inequality. The relationship between these quantities should therefore be interpreted in the context of this transformation, rather than assuming that the model applies equally across all periods. However, our analysis does not isolate the effects of institutional differences, so this interpretation remains a hypothesis for further investigation.

Another relevant assumption is that the SA model describes a closed economy. Applying it to individual countries therefore requires an approximation, given their integration through international flows of trade, capital, and income. Ellen Meiksins Wood emphasizes that national capitalist economies are interconnected within a global system governed by general capitalist laws~\cite{wood}. This perspective motivates extending the analysis beyond national boundaries as future research. 

In this sense, it would be interesting to test whether the relationship between $R$ and inequality predicted by the SA model also emerges when the world economy is treated as a single closed system. At this level, cross-border asset and liability positions would be internal to the system, removing the external-sector issue by construction. Such an analysis would nevertheless require reliable and sufficiently consistent data on the global distributions of wealth and labor compensation.
 
The empirical analysis also has measurement limitations. Definitions of income, wealth, and labor compensation differ across countries and statistical institutions, and methodologies may change over time. Constructing a common indicator such as $R$ therefore involves approximations that should be considered when interpreting the results. 

Our findings establish empirical associations rather than causal relationships. We assess whether the association predicted by the model is supported by observed wealth and income Gini coefficients; identifying its causal mechanisms would require further investigation. The results nevertheless support the usefulness of the ``Social Architecture of Capitalism'' framework, and of agent-based models more generally, for formulating testable hypotheses about inequality within capitalist economies.

The broader motivation for this investigation echoes a classical observation by Adam Smith in The Wealth of Nations: a society’s prosperity cannot be separated from the living conditions of the majority or from the share of production received by those whose labor sustains it~\cite{smith1776wealth}.


{\bf Acknowledgments:}
We acknowledge financial support from CAPES (Fundação Coordenação de Aperfeiçoamento de Pessoal de Nível Superior), through finance code 001, 
and CNPq (Conselho Nacional de Desenvolvimento Científico e Tecnológico) for Universal 406820/2025-2. 
C.A. also acknowledges partial financial support from CNPq for grant 308347/2025-0) and FAPERJ (Fundação de Amparo à Pesquisa do Estado do Rio de Janeiro), for CNE E-26/204.130/2024. S.G. would also like to thank CNPq for partially supporting this work under grant 309560/2025-0.
 
\bibliographystyle{unsrt}
\bibliography{referencias}

\end{document}